\documentstyle[12pt]{article}

\def\be{\begin{equation}}
\def\ee{\end{equation}}
\def\ba{\begin{eqnarray}}
\def\ea{\end{eqnarray}}
\def\tw{\theta_{\rm W}}

\begin{document}

\begin{titlepage}
\begin{center}

{\large \bf Topflavor: A Separate SU(2) for the Third Family}
\vskip 0.3in
{\bf David J. Muller}\footnote{e-mail address: qgd@okstate.edu} and
{\bf Satyanarayan Nandi}\footnote{e-mail address: shaown@vms.ucc.okstate.edu}\\

\vskip 0.2in

{\em Department of Physics,\\ Oklahoma State University,\\
Stillwater, OK 74078}\\

\vskip 0.2in

\end{center}

\vskip 0.6in

\begin{center}  {\bf ABSTRACT}  \end{center}
\begin{quotation} \noindent

We consider an extended electroweak gauge group: SU(2)$_1 \times$ SU(2)$_2
\times$ U(1)$_Y$ where the first and second generation of fermions couple to
SU(2)$_1$ and the third generation couples to SU(2)$_2$. Bounds based on
heavy gauge boson searches and current precision electroweak measurements
are placed on the masses of the new heavy gauge bosons. In particular we find
that the mass of the heavy W boson can not be less than 800 GeV. For some
range of the allowed parameter space, these heavy gauge bosons produce
observable signals at the Tevatron and LEP-II.

\end{quotation}

\end{titlepage}

There is no experimental data that unequivocally runs counter to
the predictions of the standard SU(2)$\times$U(1) gauge theory [\ref{a}]. It
is
reasonable, however, to ask if a model of the electroweak interaction based
on an enlarged gauge group can be constructed.
In particular, we can inquire about the possibility that
the electroweak symmetry breaks down from a higher symmetry at a scale that
will be accessible to the next generation of colliders, the 1 TeV scale, or
better yet, at present colliders such as the Tevatron or LEP II.

Perhaps the simplest extension of the standard model gauge group that one
could consider is to
include an extra SU(2). The model we propose is based on the gauge group
SU(2)$_1 \times$SU(2)$_2 \times$U(1)$_Y$ where the first and second
generation of fermions couple to SU(2)$_1$ and the third generation couples
to SU(2)$_2$.
Other models based on the gauge group
SU(2)$\times$SU(2)$\times$U(1) have been considered in the past; these include
the left-right symmetric model [\ref{e}], an un-unified model in which one
SU(2) couples
to quarks only while the other SU(2) couples to leptons only [\ref{f}], and
a model in which only one of the SU(2)'s couples to the fermions
[\ref{g}]. The particular model we propose is different from those above
and was briefly considered by Li and Ma as an approximation to their model of
generation non-universality [\ref{h}]. Moreover, this particular gauge
group has also been used in a non-commuting extended technicolor model
[\ref{cst}]. Our model is somewhat analogous to the
Topcolor model [\ref{Hill}] in that the flavor sector is extended to give
different weak interactions for the third family including the top quark;
hence the model is named Topflavor. There is also the possibility that this
new SU(2) could be used to break the Standard Model weak interaction symmetry
dynamically via the Nambu-Jona-Lasinio mechanism [\ref{njl}].

In this work we consider the phenomenology of this extended electroweak
model. We set bounds from experimental data on the masses and mixing of
the new gauge bosons of this theory. We find that the new gauge bosons can
be as light as 800 GeV. We then consider the possibility
for detecting these new gauge bosons at present and future colliders. For
some range of the SU(2)$_1$ and SU(2)$_2$ couplings, these gauge bosons
give rise to detectable physics at the Tevatron and LEP-II. We then
conclude with some remarks concerning flavor changing neutral current
interactions. The
possibilities for dynamical symmetry breaking in this model and its
implications will be discussed in a future paper.

We now give a brief overview of the model.
The quarks and the leptons of the first generation have the
following representations under ( SU(2)$_1$, SU(2)$_2$, U(1)$_Y$ )\,:
\ba
\nonumber (u, d)_L \rightarrow (2,1,1/3), & u_R
\rightarrow (1,1,4/3), \\
\nonumber d_R \rightarrow (1,1,-2/3), & (\nu_e, e)_L \rightarrow (2,1,-1), \\
e_R \rightarrow (1,1,-2) \ .
\ea
The second generation fermions have the same representations. The fermions of
the third generation have the representations:
\ba
\nonumber (t, b)_L \rightarrow (1,2,1/3), & t_R
\rightarrow (1,1,4/3), \\
\nonumber b_R \rightarrow (1,1,-2/3),
& (\nu_\tau, \tau)_L \rightarrow (1,2,-1), \\
\tau_R \rightarrow (1,1,-2) \ .
\ea
With these representations for the fermions, the theory is anomaly free.
The covariant derivative is
\be
D_\mu = \partial_\mu - i\frac{g'}{2} Y {\rm B}_\mu - ig_1 {\rm T}^a {\rm
W}^a_\mu
- ig_2 \widetilde{\rm T}^b \widetilde{\rm W}^b_\mu
\ee
where the ${\rm W}^a$ belong to ${\rm SU(2)}_1$ and the $\widetilde
{\rm W}^b$
belong to ${\rm SU(2)}_2$.

The symmetry breaking is accomplished in two steps. First the two SU(2)'s
are broken down to the ${\rm SU(2)}_{\rm W}$ of the standard model (SM). Then
the
remaining symmetry is broken down to U(1)$_{\rm em}$:
\be
{\rm SU(2)}_1 \times {\rm SU(2)}_2 \times U(1)_Y \rightarrow SU(2)_{\rm W}
\times U(1)_Y \rightarrow U(1)_{\rm em}
\ee
where the electromagnetic group is generated by
$Q = T_3 + \widetilde{T}_3 + Y/2$.

The first stage in breaking the symmetry is accomplished
by introducing a Higgs field $\Phi$ that transforms as a doublet under
each SU(2) with the vacuum expectation value (vev)
\be
\langle \Phi \rangle = \frac{1}{\sqrt{2}} \pmatrix{ u & 0 \cr 0 & u \cr } \ .
\ee
The SM symmetry group is then broken down to
${\rm U(1)_{em}}$ through the introduction of the Higgs doublet,
H = (1, 2, 1), with the vev, $\langle {\rm H} \rangle = (0, v)$. We can
generate masses for the third generation of fermions with this doublet.
We can generate masses for the first and second family by introducing
another Higgs doublet that transforms as (2, 1, 1) although we perfrom the
analysis here with just the one Higgs doublet.

The gauge bosons of the theory obtain mass through their interaction with
the Higgs fields. The mass matrix for the neutral gauge sector is
\be
\frac{1}{2} \left [ \matrix{ g_1^2 u^2 & -g_1 g_2 u^2 & 0 \cr
-g_1 g_2 u^2 & g_2^2 (v^2 + u^2) & -g' g_2 v^2 \cr
0 & - g' g_2 v^2 & g'^2 v^2 \cr } \right ]
\ee
where the basis is ordered as $W$, $\widetilde W$, $B$. This matrix can be
diagonalized by means of an orthogonal matrix which we shall call {\bf R}:
\be
\pmatrix{ W_3 \cr \widetilde{W_3} \cr B \cr } = {\rm \bf R^\dagger }
\pmatrix{ A \cr  Z_l \cr  Z_h \cr }
\ee
where the mass eigenstate are denoted by $A$, $Z_l$, and $Z_h$. The
eigenstate $A$ has zero mass and is identified as the photon. The couplings
of our theory are related to the electric charge
by
\be
g_1 = \frac{e}{\cos \phi \sin \tw} \, , \quad
g_2 = \frac{e}{\sin \phi \sin \tw} \, , \quad
g' = \frac{e}{\cos \tw}
\ee
where $\tw$ is the weak mixing angle and $\phi$ is an additional mixing
angle. We introduce the parameter
$\epsilon$ which is defined to be $v^2 / u^2$. Then the masses of the other
two mass states are given by the equation
\be
M^4_{Z_i} - \frac{1}{2} u^2 (g_1^2 + g_2^2 + g'^2 \epsilon
+ g_2^2 \epsilon) M^2_{Z_i} + \frac{1}{4} u^4 \epsilon (g_1^2 g'^2 +
g_1^2 g_2^2 + g_2^2 g'^2) = 0
\ee
where $i = l,\, h$. $Z_l$ is taken as the eigenstate with the lower
mass and is associated with the observed $Z$ boson. $Z_h$
is refered to in this paper as the ``heavy Z boson''.
For small $\epsilon$, the mixing matrix has the following approximate form:
\be
\footnotesize {\rm \bf R} = \left [ \matrix{ \cos \phi \sin \tw &
\sin \phi \sin \tw & \cos \tw \cr \cos \phi \cos \tw
+ \epsilon \frac{\cos^3 \phi \sin^2 \phi}{\cos \tw} & \sin \phi \cos \tw
- \epsilon \frac{\sin \phi \cos^4 \phi}{\cos \tw} & - \sin \tw \cr
- \sin \phi + \epsilon \sin \phi \cos^4 \phi & \cos \phi
+ \epsilon \sin^2 \phi \cos^3 \phi  & - \epsilon \tan \tw \sin \phi \cos^3
\phi \cr }
\right ]
\ee

The mass matrix for the charged sector is
\be
\frac{1}{2} \left [ \matrix{ g_1^2 u^2 & - g_1 g_2 u^2 \cr - g_1 g_2 u^2 &
g_2^2 (v^2 + u^2) \cr} \right ]
\ee
where the basis is ordered as $W$, $\widetilde W$. We denote by {\bf R}$'$ the
matrix
that diagonalizes the mass matrix:
\be
\pmatrix{ W \cr \widetilde {W} \cr } = {\rm \bf R}'^\dagger
\pmatrix{ W_l \cr W_h \cr }
\ee
where $W_l$ and $W_h$ are the mass eigenstates whose mass are obtained by
solving the equation
\be
M^4_{W_i} - \frac{1}{2} u^2 (g_1^2 + g_2^2 + g_2^2 \epsilon)
M^2_{W_i} + \frac{1}{4} u^4 g_1^2 g_2^2 \epsilon
\ee
for $i = l, h$. $W_l$ is taken as the eigenstate with the lower mass and is
associated with the observed W bosons. $W_h$ is refered to in
this paper as the ``heavy W boson''. For $\epsilon = 0$, $W_h$ and $Z_h$ are
degenerate due to a global SU(2) symmetry. For small $\epsilon$, the mixing
matrix has the approximate form
\be
{\rm \bf R}' = \left [ \matrix{ \cos \phi + \epsilon \sin^2 \phi \cos^3 \phi
& - \sin \phi + \epsilon \sin \phi \cos^4 \phi \cr
\sin \phi - \epsilon \sin \phi \cos^4 \phi
& \cos \phi + \epsilon \sin^2 \phi \cos^3 \phi \cr} \right ]  \ .
\ee
It can be easily checked that in the limit of $\epsilon = 0$, we recover the
SM couplings of all the quarks and leptons to the light gauge
bosons.

 In this work we discuss the phenomenology in which both SU(2)
interactions are perturbative.
In order for this theory to be perturbative, the parameter $\phi$ can take
only certain values. This range of values is delimited by $\tan \phi
> 0.2$ from $\frac{g_2^2}{4 \pi} < 1$ and $\tan \phi < 5.5$ from$\frac{g_1^2}{4 \pi} < 1$.

We now determine what limits we can place on the two additional parameters of
the theory, which we choose to be $M_{W_h}$ and $\phi$, from current
experimental
data. We first establish limits on the allowed values of $M_{W_h}$ and $\phi$
from precision electroweak experiments and then obtain bounds from heavy
boson searches at hadronic colliders. The main result is shown in Fig. 1.
The curve shown there cuts the $M_{W_h}$ -- $\tan \phi$ plane into an allowed
region and a disallowed region where the upper region of the plane represents
the allowed region.

We first considered bounds on the model parameters by looking at the current
precision electroweak data. The quantities considered include $R_b$, $R_c$,
$R_e$, $\rho$, $\sigma ^p_h$ which is defined as the $Z$ boson's
cross section
into hadrons at resonance, and $\frac{g_\tau}{g_\mu}$ where $g_\tau$
and $g_\mu$ are the gauge couplings of the $W_l$ to the $\tau$ and $\mu$
respectively [\ref{i}]. In our model
all of these processes have extra
contributions at tree level beyond the SM.
Thus there is the possibility that our model's theoretical values for these
quantities could significantly deviate from the current experimental values
[\ref{i}],
especially if the heavy $Z$ boson mass is in the TeV scale or less.

The most restrictive of the above quantities (the only one that contributes
to the bound that Fig. 1 represents) is $\frac{g_\tau}{g_\mu}$ which provides
a measure of the deviation from $\mu$-$\tau$ universality: the left side of
Fig. 1
was obtained by requiring that our tree level calculation for
$\frac{g_\tau}{g_\mu}$ agree with the experimental value of $1.002 \pm 0.005$
[\ref{i}] to within three standard deviations. In this calculation, the heavy
$W$ mass and $\tan \phi$ are input parameters. The result is that for a given
value of $\tan \phi$, $\frac{g_\tau}{g_\mu}$ increases (towards unity) for
increasing mass and thus
sets a lower bound on the heavy $W$ boson mass.

The right hand portion of the restriction curve is based on a heavy W boson
search that was performed by the D0 Collaboration [\ref{j}]. Their search
involved looking for evidence of an additional $W$ boson, which they call
$W'$, decaying into
an electron and its antineutrino. For various values of the $W'$
 mass, they obtained an upper bound on the cross
section times branching ratio for $W'$ divided by the
experimental result for the same process with the usual $W$ boson:
$
R \equiv  \sigma B (W' \rightarrow e \nu) /
{\sigma B (W \rightarrow e \nu)} \ .
$
The D0 collaboration considered $W'$ masses up to 800 GeV for which it
established an upper bound on $R$ of $3 \times 10^{-4}$ [\ref{j}].

In order to use their bounds on $R$ to set bounds on the heavy W boson mass
of our model, we calculate the $W_h$ production cross section at $\sqrt{s}
=$ 1.8 TeV using the CTEQ distribution. Input parameters to this calculation
include $\tan \phi$ and the heavy W mass. The couplings of the quarks to
$W_h$ are approximated to order $\epsilon$; this is satisfactory since
$\epsilon$ is small ($< 0.35$) in the region allowed by $\sigma_p^h$.
The coupling
of the first generation fermions to $W_h$ goes as $\tan^2 \phi$ so that
the $W_h$ production cross section increases as $\phi$ increases.

Next, the branching ratio $B(W_h \rightarrow e \nu)$ is calculated. As
before the input parameters to this calculation are the heavy W mass,
$M_{W_h}$,
and $\tan \phi$, but this time the couplings of the fermions to $W_h$ are
calculated to all orders in $\epsilon$. The branching ratio for $W_h$ has
additional contributions that the $W_l$ lacks from the tri-gauge
boson process $W_h \rightarrow W_l Z_l$. Superficially, one might expect
 this process to dominate over the other contributions to the branching
ratio since $\Gamma (W_h \rightarrow W_l Z_l) \sim M_h^5/M^4_l$ for
$M_h >> M_l$. However, to zeroth order in $\epsilon$ the coupling of the
above gauge bosons vanishes and, as a consequence, the contribution of this
process to the branching fraction is rather small for $M_{W_h}$ up to the few
TeV
range. For $M_{W_h}$ around a few TeV, $B (W_h \rightarrow e \nu)$ ranges
from 8 to 12.5\% in the allowed range for $\tan \phi$.

 Finally, we obtain our value for $R$ by taking the  product of the cross
section and branching ratio and then dividing by the experimental value for
$\sigma B(W \rightarrow e \nu)$ of $2.36 \pm 0.07$ nb (the value quoted by the
CDF Collaboration is $2.49 \pm 0.12$) [\ref{k}]. At a $W_h$
mass of 800 GeV, our values for $R$ are greater than $6 \times 10^{-4}$ for
$\tan \phi > 0.8$. This is greater than the bound of $3 \times 10^{-4}$
set by
the D0 Collaboration and the $R$ value gets larger with increasing $\tan
\phi$. Moreover, the restriction curve from $\frac{g_\tau}{g_\mu}$ falls to
800 GeV
around $\tan \phi = 2.1$. Therefore, we can rule out any mass for $W_h$ below
800 GeV in this model as shown in Fig. 1. This is consistent with a CDF
lower bound on the mass of a new heavy W boson of  750 GeV [\ref{l}].
Anticipating that
the bound on $R$ for the Tevatron will improve as more data is accumulated,
we show in Fig. 2 the values for $R$ that are
obtained at the Tevatron for values of $M_{W_h}$ at 850, 900, 950, and 1000
GeV.

The $\rho$ parameter in this model, $\rho = M^2_{W_l}/M^2_{Z_l} \cos^2 \tw$,
is one
to zeroth order in $\epsilon$. To all orders in $\epsilon$, the constraints
from the $\rho$ parameter as well as $K^\circ - \overline{K}^\circ$ and
$B^\circ - \overline{B}^\circ$ mixing do not restrict the allowed region
further. The $\tau$ lifetime as calculated in this model is within three
standard deviations of the
experimental value of $\tau_\tau = 289.2 \pm 1.7$\,fs
for $\alpha_s (m_\tau)$ in the range 0.32 to 0.38.

Next we consider the possibility for detecting the new heavy gauge bosons at
the future leptonic colliders.
At LEP-II the process $e^+ e^- \rightarrow W_l^+ W_l^-$ will be an important
test of
the SM prediction for the coupling of the three gauge boson
coupling of the $Z$ to the $W$. In our model it is possible to have deviations
from the standard model prediction for the cross section of this process
due to extra
terms from heavy $Z$ boson exchange as well as due to the deviation of the
couplings from the SM values.
Unfortunately, the $W_l$ pair production cross section at $\sqrt{s} =$ 200 GeV
is not significantly different from the SM value for $Z_h$ in the
few TeV range. This is largely due to the fact that the coupling
of $W_l$ to $Z_h$ vanishes to zeroth order in $\epsilon$ (similar to the
coupling of $W_h$ to $W_l$ and $Z_l$ as mentioned above).

A process whose cross section at LEP-II can be significantly different from
the
SM value is $e^+ e^- \rightarrow \mu^+ \mu^-$ since the coupling
of the electron and muon to the heavy Z boson goes as $\tan \phi$. The
values of the cross sections at $\sqrt{s} =$ 180 and 200 GeV as functions of
$\tan \phi$ are shown in Fig. 3. As $\tan \phi$ increases from 1 to 5.5, the
cross section increases from close to the SM value to about twice that value.
Thus, measurements of this cross section at LEP-II will either detect the
effect of $Z_h$ or eliminate a substantial part of the allowed region in
Fig. 1.

We now briefly state our results for future hadronic colliders. At an upgraded
Tevatron with
$\sqrt{s} =$ 4.0 TeV, the production cross section for a 1 TeV $W_h$
increases from about 1 nb at $\tan \phi = 1$ to 27 nb at $\tan \phi = 5.5$.
At the LHC with a $\sqrt{s} =$ 14 TeV, the cross section for a 1 TeV $W_h$
increases from 11 nb at $\tan \phi = 1$ to 330 nb at $\tan \phi = 5.5$ while,
for a 3 TeV $W_h$, the corresponding values are 0.6 and 19 nb. Thus, if the
heavy gauge bosons are in these mass regions, they will be within the
discovery reach of these colliders.

If the heavy $W$ mass is in the TeV or less range, the cross section for
single top
production at the LHC ($\sqrt{s} = 14$\,TeV) through the heavy $W$ boson
resonance
can be comparable to or even
dominate over that for top pair production via strong interaction processes.
The values we obtained for the cross section
are shown in Fig. 4. We see that, for example, a 1 TeV heavy $W$ with $\tan
\phi = 1.6$ gives a cross section for single top production through the
heavy W resonance of 1 nb which is approximately the cross section for top
production through gluon fusion. An interesting signal for this single top
production is two jets with a high $p_T$ lepton. Here both jets are
b-quark jets and can be tagged. This signal is different from that given by
the usual
$t \overline{t}$ production of the Standard Model.

Since the coupling for the third family ($g_2$) differs from that of the
first and second family ($g_1$), this model gives rise to flavor changing
neutral current (FCNC) interactions. The off-diagonal parts of the FCNC
Lagrangian can be written as
\ba \label{fcnc}
\nonumber {\cal L}^{off-diag}_{neutral} = \overline{U}_L \gamma^\mu
X_L^\dagger
[ d_l^{(u)} Z_{l,\mu}  +  d_h^{(u)} Z_{h,\mu} ] X_L U_L
\\
+ \ \overline{D}_L \gamma^\mu Y_l^\dagger [ d_l^{(d)} Z_{l,\mu}  +
d_h^{(d)} Z_{h,\mu} ] Y_L D_L
\ea
where $U_L \equiv (u, c, t)_L$ and $D_L \equiv (d, s, b)_L$. Only the 3-3
elements of the four
matrices $d_l^{(u)}$, $d_h^{(u)}$, $d_l^{(d)}$ and
$d_h^{(d)}$ are nonzero and these are expressed in terms
of
$g$, $M_{W_h}$ and $\phi$. $X_L$ and $Y_L$ are defined by the following
biunitary transformations:
\ba
\nonumber X_L^\dagger M_u X_R = (M_u)_{diag}  \\
Y_L^\dagger M_d Y_R = (M_d)_{diag} \ .
\ea
$X_L$ and $Y_L$ are related by the CKM matrix, $K$, by $K = X_L^\dagger Y_L$.
Thus we can use the CKM matrix to eliminate $X_l$ from eq.\,[\ref{fcnc}] to
obtain
\ba \label{fcnc2}
\nonumber {\cal L}^{off-diag}_{neutral} = \overline{U}_L \gamma^\mu K
Y_L^\dagger
[ d_l^{(u)} Z_{l,\mu} + d_h^{(u)} Z_{h,\mu} ] Y_L K^\dagger U_L \\
+ \ \overline{D}_L \gamma^\mu Y_l^\dagger [ d_l^{(d)} Z_{l,\mu} +
d_h^{(d)} Z_{h,\mu} ] Y_L D_L  \ .
\ea
Since the mass matrices are not known, we can not calculate $Y_L$ appearing
in eq.\,[\ref{fcnc2}].
However, using the observed hierarchy of the CKM matrix,
it is easily seen that the FCNC couplings in the up and down sectors are
related. For example, using $K_{ii} \simeq 1 >> K_{ij}$ for $i \not = j$ and
so neglecting the terms involving $K_{ij}$, the 2-3 elements of the FCNC
involving $Z_l$ can be written as
\ba
\nonumber (2-3)_{up} \simeq \{ \overline{c}_L [ (d^{(u)}_l)_{33}
(Y_L^\dagger)_{23} (Y_L)_{33} ] t_L + h.c. \} Z_l   \\
(2-3)_{down} \simeq \{ \overline{s}_L [ (d^{(d)}_l)_{33}
(Y_L^\dagger)_{23}
(Y_L)_{33} ] b_L + h.c. \} Z_l \ .
\ea
We get similar expressions involving $Z_h$. Thus, the FCNC interactions
involving $b \overline{s}$ will constrain those in the $t \overline{c}$
sector and vice-versa.

\pagebreak

\begin{center}
{\Large \bf Figure Captions}
\end{center}

\begin{description}

\item[Fig. 1.] Bounds on the heavy W mass (in GeV) as a function of $\tan
\phi$. The region below the curve is excluded. For $\tan \phi < 2.1$,
the bounds were obtained from $\frac{g_\tau}{g_\mu}$, while the bounds for
$ \tan \phi
\geq 2.1$ were obtained from the limits on $R$ set by the D0 Collaboration.

\item[Fig. 2.] The values of $R$ as a function of $\tan \phi$ for the Tevatron
collider ($\sqrt{s} =$ 1.8 TeV).

\item[Fig. 3.] The cross sections (in pb) for $e^+ e^- \rightarrow \mu^+
\mu^-$ as functions of $\tan \phi$ for $\sqrt{s} =$ 180 and 200 GeV.
The SM cross sections are 3.8 and 3.0 pb respectively. For
each energy the four curves correspond to $M_{Z_h}$ = 850, 900, 950, and
1,000 GeV.

\item[Fig. 4.] The cross sections (in nb) for single top production at
the LHC ($\sqrt{s} = 14$\,TeV). The six curves correspond to $M_{W_h} =$
850, 900, 950, 1000, 2000, 3000 GeV.

\end{description}

\end{document}